\documentclass[12pt]{article}

\usepackage{times}

\usepackage{amsmath,amssymb,amsfonts} 
\usepackage{subeqnarray}
\usepackage{graphicx}                 
\usepackage{color}                    
\usepackage{hyperref}                 
\usepackage[numbers]{natbib}
\usepackage{makeidx}

\DeclareGraphicsExtensions{.jpg}

\makeindex

{\end{quote}}

\title{A study of pendular liquid bridge between two equal solid spheres} 

\author
{James Q. Feng\\
\\
\normalsize{OXCBO Research, Maple Grove, Minnesota, USA}\\
\\
\normalsize{E-mail:  james.q.feng@gmail.com}
}

\date{}

\begin{document} 


\baselineskip24pt


\maketitle 


\begin{abstract}
\noindent Pendular liquid bridges with concave meridians between two equal rigid spheres are mathematically studied emphasizing some less analyzed facts in the literature.  Discrepancies from the numerical solution of the Young-Laplace equation are examined among typical simplifying approximations and a few curve-fitting formulas.  An in-depth analysis is provided about the important role of separation distance between spheres played in pendular ring formation via capillary condensation at a given relative humidity and the strength of subsequent capillary forces.  For most practical situations, the toroidal approximation could be reasonably accurate (especially with diminishing separation distance) and provide valuable mathematical insights at least in a qualitative sense with its relatively simple analytical formulas.  Using the elliptic meridional profile generally offers more accurate approximations, but with such complicated analytical formulas that it would limit its convenience for practical applications.  With a few examples, the present study shows that curve-fitting formulas cannot be perfect and, by their approximative nature, would always leave room for improvements; therefore, care should be taken when applying curve-fitting formulas, to avoid undesirable errors.   
\end{abstract}


\section{Introduction}
Everyday experience indicates that moist sand (as well as soils, powders, and granular materials in general) can be easily shaped for building castles, while dry sand can flow without much cohesivity, yet too much moisture would make the sand soggy.  Early studies of soil cohesion property have suggested the significant effects of capillary forces when wet \cite{haines25, haines27}.  To enable quantified analysis, the “ideal soil” with uniform rigid spheres in regular packing was considered to understand the moisture effects on soil mechanics.  With an idealized model, it is shown that the presence of a small amount of liquid between two solid spheres would form a so-called pendular ring meniscus which can exert substantial capillary forces on the spheres.  Even when these spheres are not in solid-solid contact, they could still be connected by a liquid bridge with the pendular ring.  The capillary forces associated with such an axisymmetric meniscus of the pendular liquid bridge have been the subject of serious investigation by numerous authors over a century \cite{fisher26}--\cite{kruyt17}, due to their relevance to many natural phenomena and practical processes. Quantitative calculations of the capillary forces due to a pendular ring meniscus can be important for understanding not only the adhesion-cohesion of powder particles, but also the behavior of insects, frogs, and geckos \cite{huber05, qian06}, as well as sintering of ceramic and metallic particles \cite{hwang87, anestiev99}, even relevant to stiction of slider in magnetic memory hard discs \cite{mate92}.
    
The physical mechanism for forming a liquid bridge may be described by the Kelvin equation which indicates that capillary condensation could occur on a liquid surface with negative mean curvature even in a subsaturated vapor environment.  Moreover, the excess pressure due to the mean curvature of the liquid bridge surface is a significant contributor to the capillary adhesion force between particles. Therefore, efforts have been made to determine the geometric shape and the associated value of mean curvature of a liquid bridge surface, as the solution of the well-established Young-Laplace equation which can only be computed numerically \cite{orr75, lian93}.  Studies with the toroidal approximation \cite{fisher26, megias09, megias10} considerably simplified the calculations, but have been found inaccurate for some situations \cite{mcfarlane50, mason65, butt09}.  Given the inconvenience with numerical computations, Lian and Seville \cite{lian16} started to develop analytical formulas by curve-fitting numerically computed results, for general usage.  However, the curve-fitting approach could not satisfy all the needs arising from various applications; hence similar efforts have been continuing to the present day  \cite{argilaga23, bagheri24}.  Recently, Kruyt and Millet \cite{kruyt17} have suggested an interesting analytical model with an elliptic approximation of the meniscus meridional profile, significantly improving the accuracy of the toroidal model but involving lengthy algebraic manipulations. 

The present work would examine discrepancies among typical simplifying approximations and the accuracy of a few curve-fitting formulas, emphasizing mathematical explanations of some less analyzed aspects in the literature with the hope of providing enhanced understanding and confidence in calculated values. 

\begin{figure}[ht] \label{fig1}
\includegraphics[clip=true,scale=0.70,viewport=30 80 650 480]{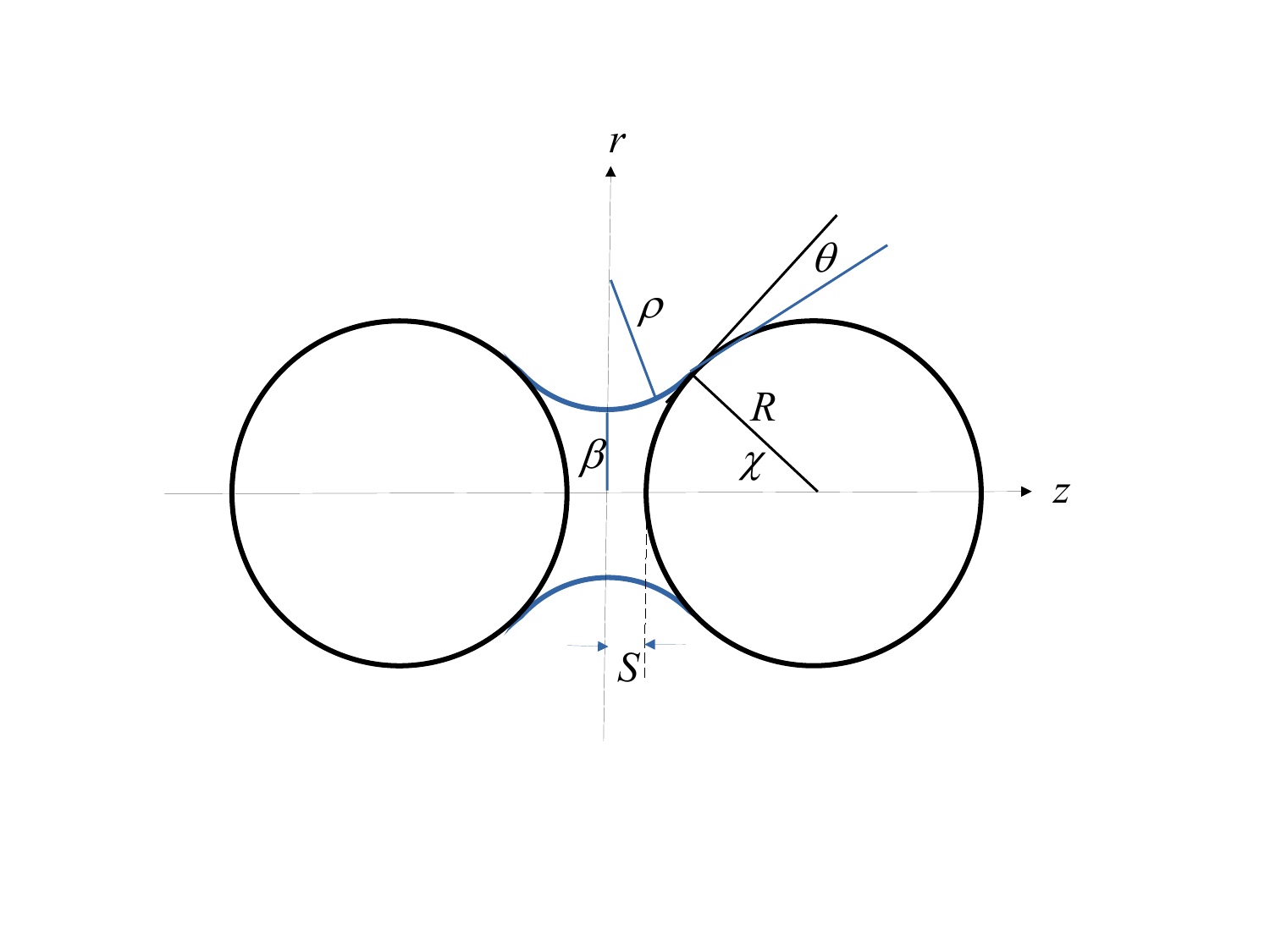}
\caption{Illustrative sketch of a pendular liquid bridge between two equal solid spheres with a half separation distance $S$ , the three-phase contact point at $z_c = S + R (1 - \cos \chi)$ and $r_c = R \sin \chi$ , liquid-solid wetting angle $\theta$ and liquid half filling angle $\chi$ (where $\beta$ denotes the bridge neck radius and $\rho$ the radius of curvature of bridge meridional profile) }
\end{figure}

\section{Problem Formulation}

Considered here is a pendular liquid bridge with a concave meridian formed between two equal solid spheres of radius $R$ separated by a distance of $2 S$, with the liquid-solid wetting angle $\theta$ and half-filling angle $\chi$ as depicted in figure 1 (based on the toroidal approximation).  Motivated by the interest in water vapor-assisted sintering of metal nanoparticle inks for printed electronics \cite{bourassa19}, the length scales in this study are assumed sufficiently small that the effect of gravity is negligible; hence, the pressure inside liquid has a constant value determined solely by the mean curvature of the axisymmetric meniscus and liquid surface tension $\sigma$.  Therefore, the free surface shape of a liquid bridge becomes the essential basis for subsequent analyses, 
and has a special place, as the case of (static) mechanical equilibrium, in studying the broader dynamic behavior of liquid bridges
\cite{liang14}-\cite{vallone23}.  

For clarity in presentation with consistent nomenclature, most of the mathematical derivations are displayed in the following subsections.  To avoid unnecessary distractions, the attention here is focused on the situation of the concave meridian, i.e., with $\chi + \theta \le \pi / 2$ and hence $\beta \le r_c$.

\subsection{Governing equations}
The shape of axisymmetric free surface r(z) is governed by the Young-Laplace equation \cite{orr75, lian93, kruyt17} 
\begin{equation}
  {2 H} = {\nabla \cdot {\bf {n}}} 
  = \frac{1}{r \sqrt{1 + (dr / dz)^2}} 
  - \frac{d^2 r / dz^2}{[1 + (dr / dz)^2]^{3/2}}
  = \frac{\Delta p}{\sigma}
  \quad  ,
  \label{Y-L}
\end{equation}	
where the unit normal vector at liquid bridge surface is defined as
\begin{equation}
  {\bf {n}} = \frac{{\bf {e_r}} - (dr / dz) {\bf {e_z}}}{\sqrt{1 + (dr / dz)^2}} 
  = \frac{(dz / dr) {\bf {e_r}} - {\bf {e_z}}}{\sqrt{1 + (dz / dr)^2}} 
  \quad  ,
  \label{normalv}
\end{equation}
and $\Delta p$ denotes the excess pressure inside liquid bridge, $\sigma$ the surface tension.

To complete the problem description, there are boundary conditions to be satisfied
\begin{equation}
  \frac{d r}{d z} = 0 
  \quad \mathrm{ at } \, z = 0 \, \, (\mathrm{ where } \, r = \beta ) 
  \quad ;
  \label{z0_bc}
\end{equation}
\begin{equation}
  \frac{d r}{d z} = \tan\left( \frac{\pi}{2} - \chi - \theta \right) 
  \quad \mathrm{ at } \, z_c = S + R (1 - \cos \chi) \, \, (\mathrm{ and } \, r_c = R \sin \chi \, ) 
  \quad . 
  \label{zc_bc}
\end{equation}
Thus, the mathematical problem becomes well-defined with the specified values of $R$, $S$, $\chi$, and $\theta$, leaving $\beta$ as a parameter determined as part of the solution.  It can also be solved by specifying $R$, $\beta$, $\chi$, and $\theta$, but leaving $S$ as a parameter as part of the solution \cite{kruyt17}, among other possibilities. 

\subsection{Solution to the Young-Laplace equation}
Now, for a rigorous calculation, the Young-Laplace equation (1) can be rearranged as
\begin{equation}
  {2 H} = {\nabla \cdot {\bf {n}}} 
  = \frac{dz / dr}{r \sqrt{1 + (dz / dr)^2}} 
  + \frac{d^2 z / dr^2}{[1 + (dz / dr)^2]^{3/2}}
  = \frac{\Delta p}{\sigma}
  \label{Y-L2}
\end{equation}
where
\[
    \frac{d^2 z / dr^2}{[1 + (dz / dr)^2]^{3/2}}
  = \frac{d}{dr} \left[ \frac{dz / dr}{\sqrt{1 + (dz / dr)^2}} \right]
  = \frac{du}{dr} \quad  ,
\]	 
with $u = \cos \varpi$  and $\varpi$  being the angle between the tangent to the meridian curve and z-axis. Thus, the Young-Laplace equation becomes
\begin{equation}
    \frac{du}{dr} + \frac{u}{r} = 2 H
  = \frac{\Delta p}{\sigma} \quad , \mathrm{ namely } \, , \, \, 
  r \frac{du}{dr} + u = 2 H r \quad  .
  \label{Y-L3}
\end{equation}	 

The first integral of (\ref{Y-L3}) yields
\begin{equation}
   r u = H r^2 + C
   \quad  .
  \label{int1stY-L3}
\end{equation}	

At $z_c = S + R (1 - \cos \chi)$ and $r_c = R \sin \chi$ , we have $dz / dr = \tan (\chi + \theta)$ and $u = \sin (\chi + \theta)$  (where $\varpi = \pi / 2 - \chi - \theta$), yielding
\[
  C = R \sin \chi [\sin (\chi + \theta) - H R \sin \chi ]
  \quad  .
\]	 	
And as $z \to 0$, $dz / dr \to \infty$, $u \to 1$ and $r \to \beta$, leading to	
\[
  C = \beta - H {\beta}^2
  \quad  .
\]	
Therefore, either the value of mean curvature $H$ can be determined from a specified neck radius $\beta$ (as adopted in the present work) or $\beta$ from a given $H$ 
\begin{subeqnarray}
  \label{Hofbeta}
   H(\beta) = \frac{R \sin \chi \, \sin(\chi + \theta) - \beta}{(R \sin \chi + \beta)(R \sin \chi - \beta)} \quad ,  \slabel{Hofbeta0}  \\ 
  \beta(H) = \frac{1 - \sqrt{1 - 4 H R \sin \chi [\sin (\chi + \theta) - H R \sin \chi ]}}{2H}  \slabel{Hofbeta1} 
   \quad  .
\end{subeqnarray}	

Now, without completely solving the problem, it is already clear 
from (\ref{Hofbeta0}) that negative 
$H$ can be obtained when 
$\beta  > R \sin \chi \, \sin(\chi + \theta)$ 
because $R \sin \chi = r_c > \beta$ 
for a free surface with concave meridian .  
As $\chi + \theta \to \pi / 2$, we would have $\beta \to R \sin \chi$ and $2 H \to 1 / \beta$ with the meniscus shape approaching that of a cylinder.  
Then negative $H$ can only occur when the value of 
$\chi + \theta$ is substantially smaller than $\pi / 2$.  Thus, there must exist some value of 
$\chi + \theta  = \arcsin [\beta /(R \sin \chi)]$ 
in between $0$ and $\pi / 2$ for $H = 0$.  
	
The free surface shape of liquid bridge can thus be determined based on (\ref{int1stY-L3}) by integrating (as discussed in Kruyt and Millet \cite{kruyt17})
\begin{equation}
   \frac{d z}{d r} = \frac{H r^2 + C}{\sqrt{r^2 - (H r^2 + C)^2}}
   \quad \mathrm{ thus } \, \, \,
   z(r) = z_c - \int_r^{r_c} {\frac{H r^2 + C}{\sqrt{r^2 - (H r^2 + C)^2}} dr}
   \quad  .
  \label{zofr}
\end{equation}	
A physically meaningful value of $\beta$ and then $H(\beta)$ in (\ref{Hofbeta0}) can be determined iteratively via numerical computing to satisfy the boundary condition of $z(0) = 0$ at $r = \beta$ , i.e.,
\begin{equation}
   \int_{\beta}^{r_c} {\frac{H r^2 + C}{\sqrt{r^2 - (H r^2 + C)^2}} dr} = z_c = S + R (1 - \cos \chi)
   \quad  .
  \label{betaofzc}
\end{equation}	

With the value of $\beta$ and thus $H(\beta)$ in 
(\ref{Hofbeta0}) being determined, 
the value of $\Delta p / \sigma  = 2 H$ from (\ref{Y-L}) enables calculating the capillary forces on each sphere in a normalized form as (attractive if positive, following Orr et al. \cite{orr75} , Lian et al. \cite{lian93} , Kruyt and Millet \cite{kruyt17}
\begin{equation}
   F = \frac{F_{cap}}{\pi \sigma R}
   = \frac{\beta}{R} \left[ 2 - \frac{\beta}{R} \left( \frac{R \Delta p}{\sigma} \right)  \right]
   = \sin \chi \left[ 2 \sin (\chi + \theta) - \sin \chi \left( \frac{R \Delta p}{\sigma} \right) \right]
   \quad  .
  \label{Fcap}
\end{equation}
as well as the liquid bridge volume
\[
  V_{bridge} + \frac{2 \pi R^3}{3} (1 - \cos \chi )^2 (2 +\cos \chi) 
  = 2 \pi \int_{\beta}^{r_c} {\frac{r^2 (H r^2 + C)}{\sqrt{r^2 - (H r^2 + C)^2}} dr}
  \quad  , 
\]	
which leads to the normalized volume measured in units of the sphere volume
\begin{equation}
   V = \frac{3 V_{bridge}}{4 \pi R^3}
   = \frac{3}{2} \int_{\beta}^{r_c} {\frac{r^2 (H r^2 + C)}{\sqrt{r^2 - (H r^2 + C)^2}} dr}
   - \frac{(1 - \cos \chi )^2 (2 +\cos \chi)}{2}
   \quad  .
  \label{Vbridge}
\end{equation}
where the last term comes from the spherical cap of polar angle $\chi$ .

It is noteworthy that (\ref{Fcap}) indicates that the capillary forces $F > 0$ are guaranteed even when the normalized excess pressure $R \Delta p / \sigma = 2 H R = R / \beta > 0$ at its maximum which makes $F = \beta / R$.  The capillary forces in (\ref{Fcap}) can be calculated either by considering a force balance at the neck of liquid bridge $z = 0$ (e.g., Lian et al. \cite{lian93}) or at $z = z_c$ (e.g., Orr et al. \cite{orr75}), both yielding the same result as long as (\ref{Hofbeta0}) holds, consistent with physical expectation. 

\subsection{Toroidal approximation}
Over a century of study, there have not been general closed-form analytical solutions found to (\ref{Y-L})-(\ref{zc_bc}); the exact solution could only be computed numerically.  Now, if the free surface meridional profile $r(z)$ is approximated as a part of the circle of constant radius $\rho$, i.e., with a toroidal free surface \cite{fisher26, megias09, megias10},  
\begin{equation}
   z^2 + (r - \beta -\rho)^2 = \rho^2 \quad \mathrm{ or } \, \, \,
   r(z) = \beta + \rho - \sqrt{\rho^2 - z^2}
   \quad  ,
  \label{toroidalrofz}
\end{equation}
$\beta$ and $\rho$ can be explicitly determined from boundary conditions (\ref{z0_bc}) and (\ref{zc_bc}) based on the geometry. Thus, we could have (as in Lian et al. \cite{lian93}) 
\begin{equation}
   \frac{\rho}{R} = \frac{1 - \cos \chi + S/R}{\cos (\chi + \theta)} \quad \mathrm{ and } \, \, \,
   \frac{\beta}{R} = \sin \chi - \frac{\rho}{R} [1 - \sin (\chi + \theta)]
   \quad  ,
  \label{toroidalrhobeta}
\end{equation}
which leads to the curvature terms in (\ref{Y-L})
\[
   \frac{1}{r \sqrt{1 + (dr/dz)^2}} = \frac{1 - (z/\rho)^2}{r} \quad \mathrm{ and } \, \, \,
   \frac{d^2r/dz^2}{[1 + (dr/dz)^2]^{3/2}} = \frac{1}{\rho}
   \quad  .
\] 
At $z = 0$, $d r / d z = 0$ and $r = \beta$  we get 
\[
   \frac{1}{r \sqrt{1 + (dr/dz)^2}} = \frac{1 - (z/\rho)^2}{r} = \frac{1}{\beta} 
   \quad  .
\] 
But at the contact circle $z = z_c = S + R (1 – \cos \chi )$ and $r = r_c = R \sin \chi $  we get
\[
   \frac{z_c}{\rho} = \cos (\chi + \theta) \, \, , \, \,
   \frac{1}{r \sqrt{1 + (dr/dz)^2}} = \frac{1 - (z/\rho)^2}{r} = \frac{\sin (\chi + \theta)}{R \sin \chi} \le \frac{1}{\beta}  
   \quad  ,
\] 
because $R \sin \chi \ge \beta$ and $\sin (\chi + \theta) \le 1$ . 

As pointed out by Lian et al. \cite{lian93}, the toroidal free surface does not have a constant mean curvature $H$ in the Young-Laplace equation (\ref{Y-L}); rather, its free surface shape corresponds to a variable mean curvature
\begin{equation}
  {\widetilde H}_c = \frac{1}{2} 
 \left[ \frac{\sin (\chi + \theta)}{R \sin \chi} - \frac{1}{\rho} \right] 
  \le {\widetilde H} = \frac{\sqrt{1 - (z / \rho)^2}}{r} - \frac{1}{\rho}
  \le {\widetilde H}_0 = \frac{1}{2} \left( \frac{1}{\beta} - \frac{1}{\rho} \right)
   \quad  .
  \label{toroidalHrange}
\end{equation}
It may then be natural to take an average 
\begin{equation}
  {\widetilde H}_a = \frac{{\widetilde H}_0 + {\widetilde H}_c}{2}
  = \frac{3 \beta R \sin \chi [\sin (\chi + \theta) - 1] + (R \sin \chi )^2 - \beta^2 \sin (\chi + \theta)}{4 \beta R \sin \chi (R \sin \chi - \beta)}   
   \quad  ,
  \label{toroidalHa}
\end{equation}
for approximating the actual $H$.  But when it comes to calculating the capillary forces of (\ref{Fcap}), with the toroidal approximation, Lian et al. \cite{lian93} considered two formulas based on either the local force balance at the neck of liquid bridge $z = 0$ with ${\widetilde H}_0$ (as in Fisher \cite{fisher26}) or at the contact circle $z = z_c$ with  ${\widetilde H}_c$ (as in Orr et al. \cite{orr75}).  However, the calculated value differs depending on which formula is used.  In fact, as shown in (\ref{Fcap}), the two formulas can yield the same result only when ${\widetilde H}(\beta)$ is calculated via (\ref{Hofbeta0}).  
Therefore,	
\begin{equation}
  {\widetilde F}= {2 \beta (1 - \beta {\widetilde H}_{\beta})} / {R} = 2 \sin \chi [\sin (\chi + \theta) - R \sin \chi {\widetilde H}_{\beta}]   
   \quad  ,
  \label{toroidalFcap}
\end{equation}
is used here for calculating the capillary forces, with ${\widetilde H}_{\beta}$ $= {\widetilde H} (\beta)$ 
according to (\ref{Hofbeta0}) using the $\beta$ determined from (\ref{toroidalrhobeta}) for the toroidal approximation. 
This (\ref{toroidalFcap}) is a new capillary force formula with the toroidal approximation, eliminating the difference between 
the values determined according to local force balance at $z = 0$ and $z = z_c$.

With the toroidal approximation, the normalized liquid bridge volume of (\ref{Vbridge}) can now be explicitly written as
\begin{equation}
\begin{array}{lllll}
  {\widetilde V}= \frac{3}{2} \left( \frac{\rho}{R} \right)   
  \left[ 2 \left( \frac{\rho}{R} \right)^2 + 2 \frac{\rho}{R} \frac{\beta}{R} + \left( \frac{\beta}{R} \right)^2 \right] \cos (\chi + \theta) - \frac{1}{2} \left[ \frac{\rho \cos (\chi + \theta)}{R} \right]^3 - \\ \\
  - \frac{3}{2} \left( \frac{\rho}{R} \right)^2 \left( \frac{\rho}{R} + \frac{\beta}{R} \right) \left[ \cos (\chi + \theta) \sin (\chi + \theta) + \frac{\pi}{2} - (\chi + \theta) \right] - \\ \\
  - \frac{1}{2} (1 - \cos \chi)^2 (2 + \cos \chi)
\end{array}
     \quad  ,
    \label{toroidalVbridge}
\end{equation}
for straightforward calculations.

\subsection{Elliptic meridional profile}
To improve the analytical approximation, an elliptic meridional profile has been developed by Kruyt and Millet \cite{kruyt17} as 
\begin{equation}
   \frac{z^2}{a^2} + \frac{(r - m)^2}{b^2} = 1 \quad \mathrm{ or } \, \, \,
   r(z) = m - \frac{b}{a} \sqrt{a^2 - z^2}
   \quad  ,
  \label{ellipticrofz}
\end{equation}
where, based on the boundary conditions (\ref{z0_bc}) and (\ref{zc_bc}), 
\begin{equation}
\begin{array}{lllll}
   m = \frac{z_c r_c - (r_c + \beta) (r_c - \beta) \tan (\chi + \theta)}{z_c - 2 (r_c - \beta) \tan (\chi + \theta)} \\ \\
    b = m - \beta  \\ \\
   a = \frac{z_c (m - \beta)}{\sqrt{(m - \beta)^2 - (m - r_c)^2}}
\end{array}
   \quad \quad \quad .
  \label{ellipticmba}
\end{equation}

For a given half-filling angle $\chi$,  the values of $z_c$ and $r_c$ are known. When the wetting angle $\theta$ is also specified, the neck radius $\beta$ is still an adjustable parameter for determining the values of $a$, $b$, $m$ in (\ref{ellipticrofz}).

The elliptic meridional profile described by (\ref{ellipticrofz}) leads to
\[
   \frac{dr}{dz} = \left( \frac{b}{a} \right) \frac{z}{\sqrt{a^2 - z^2}} \, \, , \, \,
   \frac{d^2r}{dz^2} = \frac{a \, b}{(a^2 - z^2)^{3/2}} 
   \quad  ,
\] 
and thus,
\begin{equation}
   2 {\widetilde H}_e = \frac{[a^2 + b^2 + (b^2 - a^2) z^2/a^2] \sqrt{a^2 - z^2} - a \, b \, m}{[m - (b/a) \sqrt{a^2 - z^2}] [a^2 + (b^2 - a^2) z^2 / a^2 ]^{3/2}}
   \quad  .
  \label{ellipticHe}
\end{equation}

The closure can be brought with a requirement of the value of $\beta$ (which determines $a$, $b$, $m$) to have the same mean curvature value ${\widetilde H}_e$ at $z = 0$ and $z = z_c$ \cite{kruyt17}, namely
\begin{equation}
   \frac{a^2 + b^2 - b \, m}{a^2 \, (m - b)} = \frac{[a^2 + b^2 + (b^2 - a^2) z_c^2/a^2] \sqrt{a^2 - z_c^2} - a \, b \, m}{[m - (b/a) \sqrt{a^2 - z_c^2}] [a^2 + (b^2 - a^2) z_c^2 / a^2 ]^{3/2}}
   \quad  .
  \label{ellipticConstraint}
\end{equation}
Unfortunately, (\ref{ellipticConstraint}) is an algebraically rather complicated equation for root-finding analytically in terms of $\beta$; however, it can be computed with numerical iterations. 

\section{Results and Discussion}
For presentation generality, the data shown here are computed by specifying $R = 1$ and $\sigma = 1$ which means 
to effectively measure length in units of $R$ and excess pressure in units of $\sigma / R$.  
Hence, the values of $S$, $H$, etc. given here are the same as $S/R$, $H R$, etc., and used interchangeably.

\subsection{Case of perfect wetting ($\theta = 0$)}
With perfect wetting liquids,  we have the wetting angle $\theta = 0$ and the maximized capillary effects.  
Therefore, some authors (e.g., Erle et al. \cite{erle71}) would only focus on computing the case of $\theta = 0$.  
So, it is worthwhile taking a closer look at this special case. 

\subsubsection{Discrepancies among different approximations}
Figure 2 shows the meridional profiles computed based on the Young-Laplace equation 
along with those with the toroidal and elliptic approximations, 
for the half-filling angle $\chi = 15^o$ (and $\theta = 0$) at $S/R = 0$ ($V = 0.001197$) and $0.05$ ($V = 0.003675$).  
It reveals that the toroidal approximation can be quite reasonable when the separation distance $S/R$ is small, 
but with increasing $S/R$ the discrepancies become more obvious 
although the elliptic approximation may remain close to the Young-Laplace solution.  
  
\begin{figure}[ht] \label{fig2}
\includegraphics[clip=true,scale=0.67,viewport=30 80 720 460]{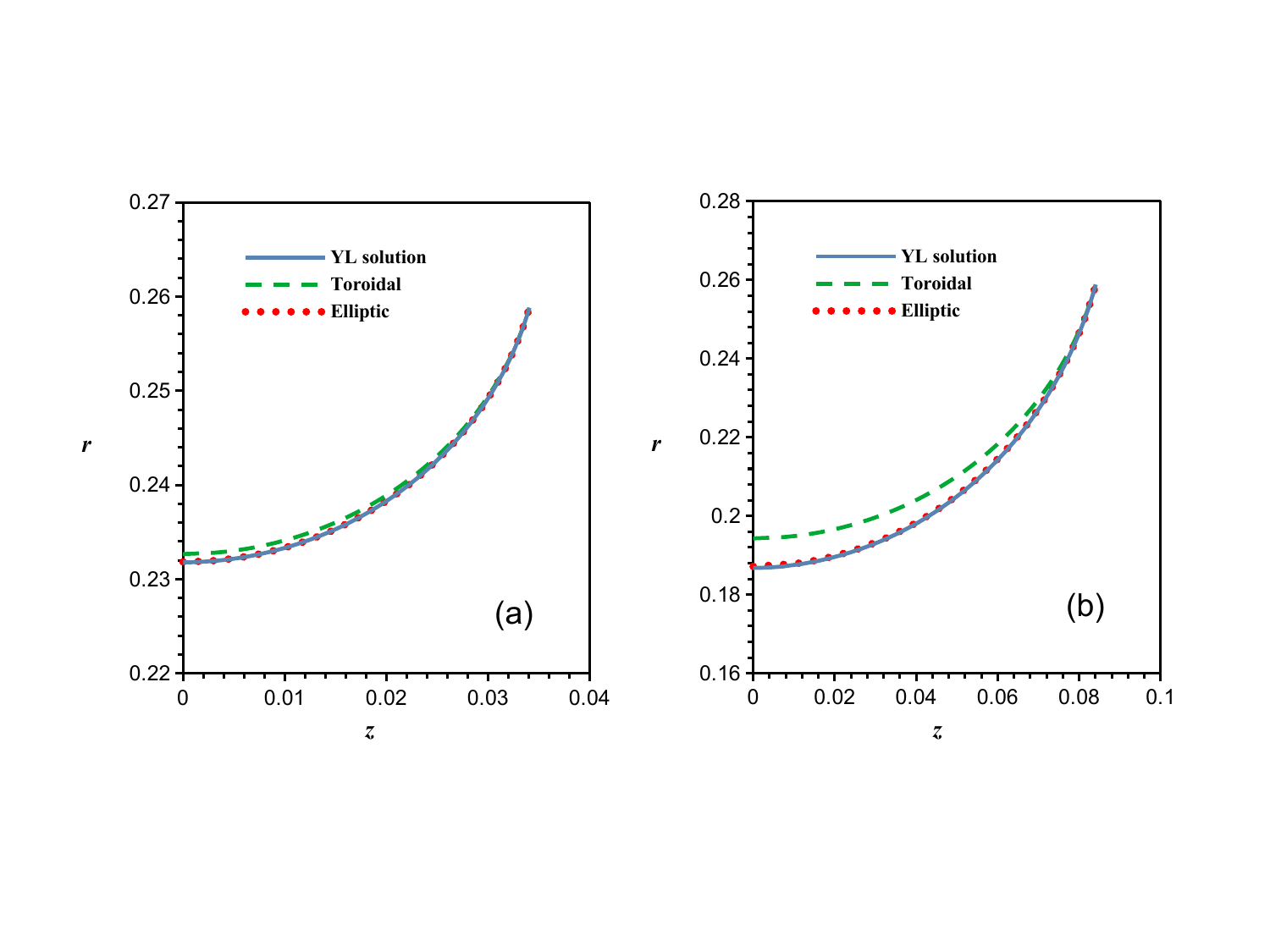}
\caption{Comparison of meniscus meridional profiles computed based on the solution of the Young-Laplace equation (\ref{zofr}), toroidal approximation (\ref{toroidalrofz}), and elliptic approximation (\ref{ellipticrofz}), for $\chi = 15^o$ and $\theta = 0$ with (a) $S/R = 0$ and (b) $S/R = 0.05$ }
\end{figure}

For a constant liquid bridge volume $V = 0.001$, figure 3 shows the meridional profiles 
at $S/R = 0.06$ ($\chi = 9.615^o$) and $0.07$ ($\chi = 10.33993^o$), indicating that 
the elliptic approximation may also deviate from the Young-Laplace solution in some circumstances.

According to the empirical formula by Lian et al. \cite{lian93} for determining the bridge rupture distance $S_c$ 
(in the present nomenclature, with $\theta$ in radian): 
\begin{equation}
   \frac{S_c(V)}{R} = \frac{2 + \theta}{4} \left( \frac{4 \pi V}{3} \right)^{1/3}
   \quad  ,
  \label{Sc}
\end{equation}
we would have $S_c / R = 0.0806$ and $0.1015$ for $V = 0.001$ and $0.002$. 
When $S / R$ is getting close to $S_c / R$,  the elliptic approximation of the meniscus shape may become less accurate 
as shown in figure 3 for the case of  $S/R = 0.07$ for $V = 0.001$, 
and similarly with  $S/R= 0.09$ for $V = 0.002$ though not shown here.  
For a given liquid bridge volume $V$, a solution for the Young-Laplace equation can only exist for $S \le S_c$ 
whereat the solution branch encounters a turning point in the numerical computation \cite{lian93}, 
as also being verified with experiments \cite{willett00}.  

\begin{figure}[ht] \label{fig3}
\includegraphics[clip=true,scale=0.67,viewport=30 80 720 460]{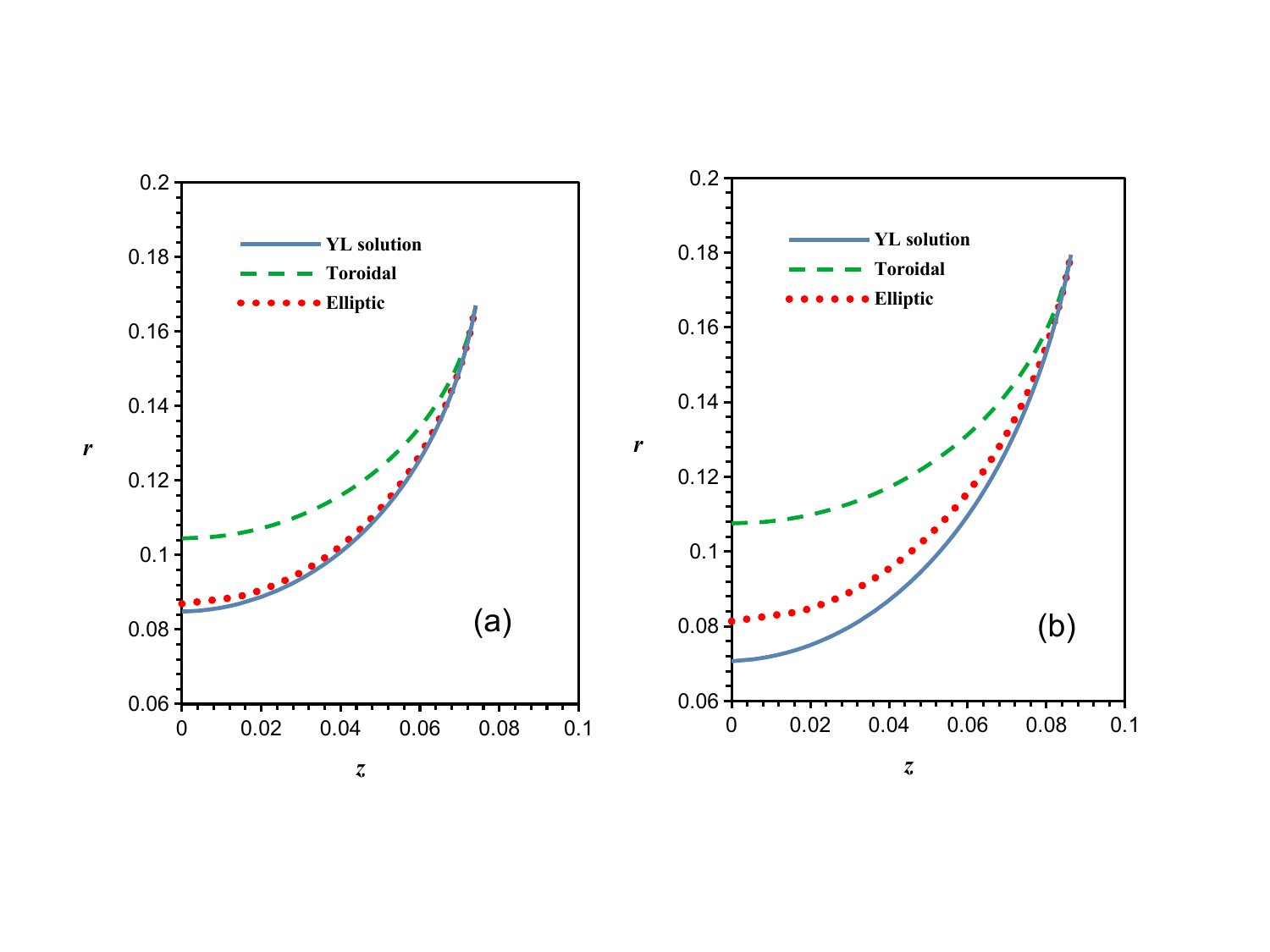}
\caption{As figure 2 but for $V = 0.001$ according to (12) with (a) $S/R = 0.06$, $\chi = 9.615^o$ , and (b) $S/R = 0.07$, $\chi = 10.33993^o$ (both at $\theta = 0$) }
\end{figure}

To provide more concrete data for convenient comparison, 
table 1 shows computed values based on the Young-Laplace equation and those from the toroidal approximation 
(${\widetilde H}_{\beta}$ in parentheses) for $\beta / R$, $H R$, and $V$  at various specified $S/R$ and $\chi$ (with $\theta = 0$).  
The comparison with the values from the Young-Laplace solutions reveals the fact that the toroidal model can be reasonably accurate 
when $\chi \to 0$  and $\chi \to \pi / 2$, 
but becomes increasingly inaccurate 
when $H$ is approaching zero, e.g., in the interval of $45^o < \chi < 65^o$ for $S/R = 0$.  
If the discrepancy of the toroidal ${\widetilde H}_{\beta}$ in parentheses from the Young-Laplace $H$ becomes too large
(e.g., $> 10\%$), 
it is natural to take a look at the toroidal 
${\widetilde H}_a$ or ${\widetilde H}_0$ as alternative choices for approximation.  
For example, at $\chi = 45^o$ and $S/R = 0$ the value ${\widetilde H}_a$ and ${\widetilde H}_0$ would be
$-0.5303$ and $-0.3536$ deviating from $H$ by $\sim 14\%$ and $24\%$, respectively (while the value of ${\widetilde H}_e$ 
with the elliptic model is $-0.4770$, much closer to the Young-Laplace $H = -0.4651$ with a discrepancy of only $2.6\%$).  
If $S/R$ is increased to $0.1$ (for $\chi = 45^o$) we would have $H = -0.1146$ but ${\widetilde H}_a = -0.1906$ and 
${\widetilde H}_0 = 0.01863$ deviating from $H$ by $66\%$ and $84\%$
(with ${\widetilde H}_{\beta} = -0.2178$), while ${\widetilde H}_e = -0.1282$ 
(again much closer to $H$ than the toroidal model, but with a discrepancy reduced to $12\%$).  
It appears that ${\widetilde H}_{\beta}$ and ${\widetilde H}_a$ follow each other rather closely, 
both differing from ${\widetilde H}_0$ quite noticeably.  
If ${\widetilde H}_{\beta}$ and ${\widetilde H}_a$ are generally comparable in terms of deviation from $H$, ${\widetilde H}_{\beta}$ 
might be preferred for consistency with the physical meaning of (\ref{toroidalFcap}).
\begin{table}
  \begin{center}
\caption{Comparison between computed values based on the Young-Laplace equation and those from the toroidal approximation (in parentheses) for $\beta / R$, $H R$, and $V$  at various specified $S/R$ and $\chi$ (with $\theta = 0$)}
  \begin{tabular}{lccccc}
  \hline
    $S / R$ & $\chi \, \mathrm{(deg)}$ & $\beta / R$ & $H R \, ({\widetilde H}_{\beta} R)$ & $V$  \\[3pt]
  \hline    
     $ $ & 5 & 0.08362 (0.08367) & -125.920 (-127.714) & $1.901 \times 10^{-5}$ ( $1.905 \times 10^{-5}$ ) \\
     $ $ & 15 & 0.2318 (0.2327) & -12.423 (-12.893) & 0.001197 (0.001212) \\
     0 & 45 & 0.5775 (0.5858) & -0.4651 (-0.5469) & 0.05278 (0.05523) \\
     $ $ & 60.63 & 0.7272 (0.7379) & 0.1397 (0.1002) & 0.1379 (0.1444) \\
     $ $ & 89.9 & 0.99903 (0.99913) & 0.49946 (0.49934) & 0.49811 (0.49826) \\
     0.05 & 11.940 & 0.1403 (0.1488) & -4.2166 (-5.1323) & 0.002001 (0.002199) \\
     0.09 & 13.2511 & 0.08583 (0.1369) & -0.7369 (-2.4942) & 0.002000 (0.003553) \\
     0.1 & 15 & 0.1123 (0.1559) & -0.8339 (-2.0847) & 0.003347 (0.005140) \\
  \hline
  \end{tabular}
  \end{center}
\end{table}

To illustrate the varying mean curvature from $z = 0$ to $z_c$ (with length in units of $R$) with the toroidal model and elliptic model, figure 4 provides two exemplifying cases: (a) $S = 0.1$ and $\chi = 45^o$ with the elliptic ${\widetilde H}_e$  being much closer to the constant $H$ than the toroidal ${\widetilde H}$; (b) $S = 0.09$ and $\chi = 13.2511^o$ with $V = 0.002$ wherewith even the elliptic ${\widetilde H}_e$ exhibits considerable deviations from the Young-Laplace solution, whereas the value of toroidal ${\widetilde H}_0$ appears as a relatively better approximation.  

\begin{figure}[ht] \label{fig4}
\includegraphics[clip=true,scale=0.67,viewport=30 80 720 440]{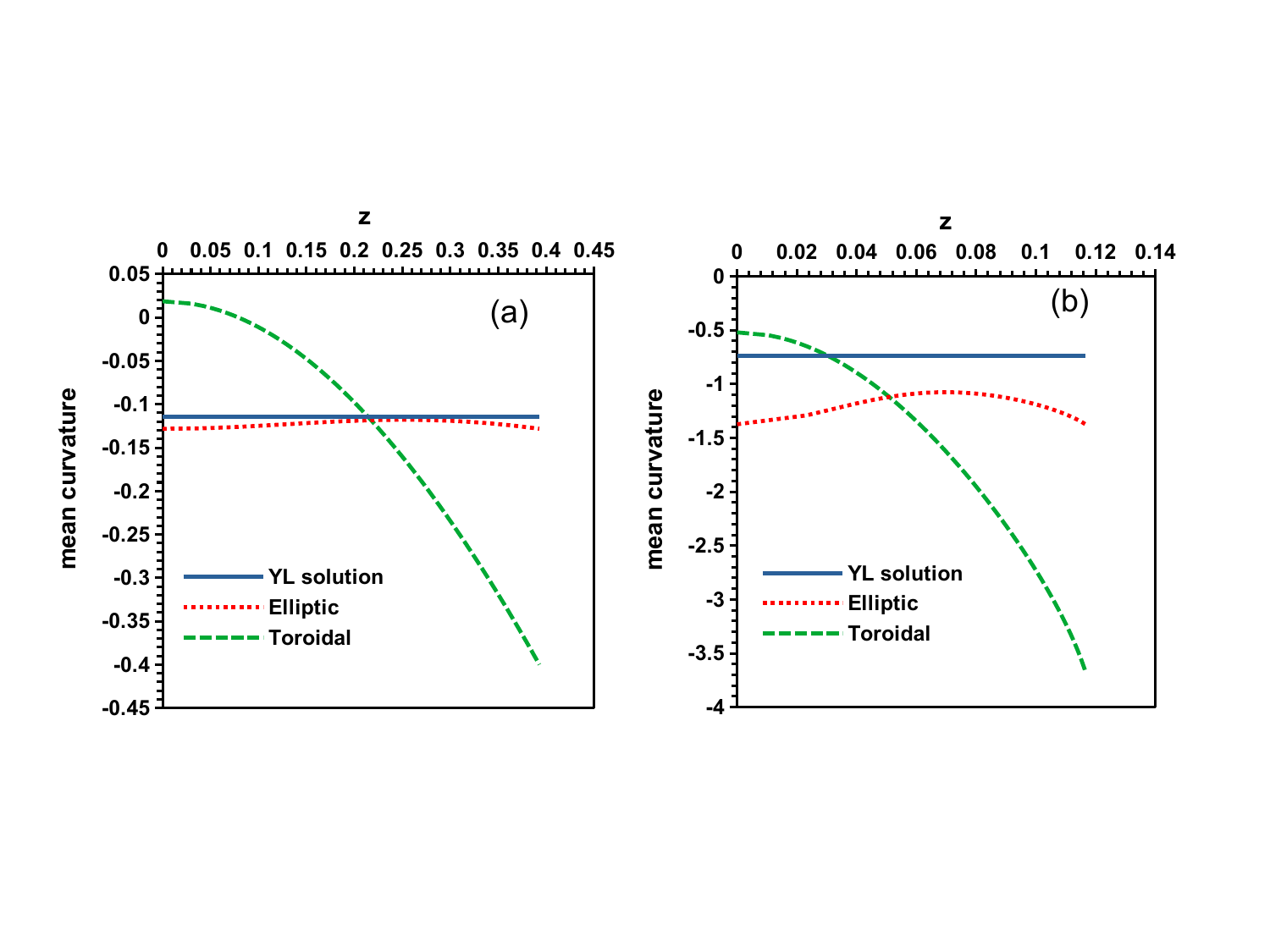}
\caption{Varying mean curvature from $z = 0$ to $z_c$ (in units of $R$) according to (15) for ${\widetilde H}$ of toroidal model and (21) of elliptic model for (a) $S/R = 0.1$, $\chi = 45^o$ with  ${\widetilde H}_{\beta} R = -0.2178$ at $z \sim 0.29$, and (b) $S/R = 0.09$, $\chi = 13.2511^o$ with $V = 0.002$, ${\widetilde H}_{\beta} R = -1.8192$ at $z \sim 0.076$ (both at $\theta = 0$) }
\end{figure}

Compared with the toroidal model, the elliptic model of Kruyt and Millet \cite{kruyt17} generally 
offers a more accurate approximation to the Young-Laplace solution for $S/R = 0$ and $\chi < 60.631^o$.  
For example, at $\chi = 60.63^o$ the values of $\beta /R$, $H R$, and $V$ from the elliptic model are $0.7281$, $0.1345$, and $0.1382$
certainly much better than those of the toroidal model shown in table 1.  However, when $\chi$ is increased beyond $60.631^o$ 
there does not seem to exist a value of $\beta /R$ that could satisfy the closure relationship (\ref{ellipticConstraint}).

With the elliptic approximation (\ref{ellipticrofz}) there actually exists a few mathematical restrictions when examining the expression (\ref{ellipticmba}) for $a > 0$, $b > 0$, and $m > 0$ with $r_c  > \beta$ to be physically meaningful.  It seems to require that
\begin{equation}
   z_c - 2 (r_c - \beta) \tan (\chi + \theta) > 0 \, \, \, \mathrm{ or } \, \,
   r_c - \frac{z_c }{2 \tan (\chi + \theta)} < \beta < r_c
   \quad  
  \label{ellipticConditions}
\end{equation}
to guarantee $m > 0$ which actually leads to $m > r_c$ .as a consequence.  Clearly, $\chi + \theta$ could not exceed a value below $\pi / 2$ due to (\ref{ellipticConditions}).  
For example, at $S = 0$ and $\chi = 60.63^o$ the lower bound for $\beta$ according to (\ref{ellipticConditions}) would be $0.7280874$, while at $\beta \sim 0.728087821$
the values of ${\widetilde H}_e$ from (\ref{ellipticHe}) 
at $z = 0$ and $z = z_c$ could match at $\sim 0.1345025$.
But when $\chi$ is increased to $60.631^o$,
even when $\beta$ approaches its lower bound value $0.728097533$ 
the values of ${\widetilde H}_e$ 
at $z = 0$ and $z = z_c$ would still be $\sim 0.134529$ and $\sim 0.134522$, could not match exactly.  
Hence, not all the values of $\beta$ that satisfy (\ref{ellipticConditions}) can also satisfy (\ref{ellipticConstraint}) for the mathematical closure.  

It is interesting to note that the toroidal approximation (\ref{toroidalrhobeta}) is equivalent to  
\[
   \beta = r_c - z_c \, \frac{1 - \sin (\chi + \theta)}{\cos (\chi + \theta)} \, \,  \left( > r_c - \frac{z_c }{2 \tan (\chi + \theta)} \right)
   \quad  ;
\]
thus, the condition (\ref{ellipticConditions}) indicates that the lower bound of the elliptic neck radius should always be smaller than the toroidal $\beta$.  As illustrated in figures 2 and 3, the toroidal neck radius appears always greater than the elliptic one which in turn (often just slightly) greater than that of the Young-Laplace solution. 

\subsubsection{Mean curvature and capillary condensation }
The phenomenon of capillary condensation is usually rendered as vapor condensation that occurs at a concave vapor-liquid interface below the saturation vapor pressure of the pure liquid.  To evaluate the mean curvature effect on vapor condensation for pendular liquid bridge formation, let’s consider the Kelvin equation 
\begin{equation}
   \frac{p_v}{p_0} =  \exp \left( \frac{2 H \sigma M_w}{\rho_l N_A k_B T} \right) \, , \, \, \mathrm{ namely } , \, \, 
   h = \exp (2 H \lambda_K) \, \, \mathrm{ or } \, \, \,
   2 H R = \frac{R \, \ln h}{\lambda_K}
   \quad  ,
  \label{kelvin}
\end{equation}
where $p_v$ and $p_0$ denote the equilibrium vapor pressure at a curved liquid surface and saturation vapor pressure 
at a liquid surface of zero curvature (namely, flat liquid surface), 
$M_w$ ($= 18.015 \times 10^{-3}$ kg mol$^{-1}$ for water) 
and $\rho_l$ ($\sim 1000$ kg m$^{-3}$ for water)
the molecular weight (or molecular mass) and density of the bulk liquid, 
$N_A$ and $k_B$ the Avogadro’s number and Boltzmann’s constant 
($N_A \times k_B  = 8.3143$ J deg$^{-1}$ mol$^{-1}$ is the so-called universal gas constant), 
and $T$ the absolute temperature.  
Also in (\ref{kelvin}) $h = p_v / p_0$ denotes the “relative humidity” 
and $\lambda_K = \sigma M_w / (\rho_l N_A k_B T)$ 
the so-called Kelvin length \cite{butt09} . 

Figure 5 illustrates how the excess pressure—related to twice the normalized mean curvature, 
i.e., $2 H R$—varies with half-filling angle $\chi$ in radians, along with discrepancies of the toroidal model 
with ${\widetilde H}_{\beta}$ in (a) and ${\widetilde H}_{0}$ in (b) in comparison with $H$ computed 
from the Young-Laplace equation. Both ${\widetilde H}_{\beta}$ and ${\widetilde H}_{0}$ deviate $H$ noticeably 
when $\chi < \pi / 4$, with ${\widetilde H}_{0}$ overestimating while ${\widetilde H}_{\beta}$ underestimating 
the magnitude of negative $H$ .  However, both ${\widetilde H}_{\beta}$ and ${\widetilde H}_{0}$ 
exhibit similar qualitative characteristics to that of $H(\chi)$: monotonic without minimum when $S = 0$ and 
having a minimum at a nonzero $\chi = \chi_{peak}$ when $S > 0$.  Such characteristics may be 
explained mathematically using the available analytical formulas with the toroidal approximation. 

\begin{figure}[ht] \label{fig5}
\includegraphics[clip=true,scale=0.65,viewport=10 80 720 440]{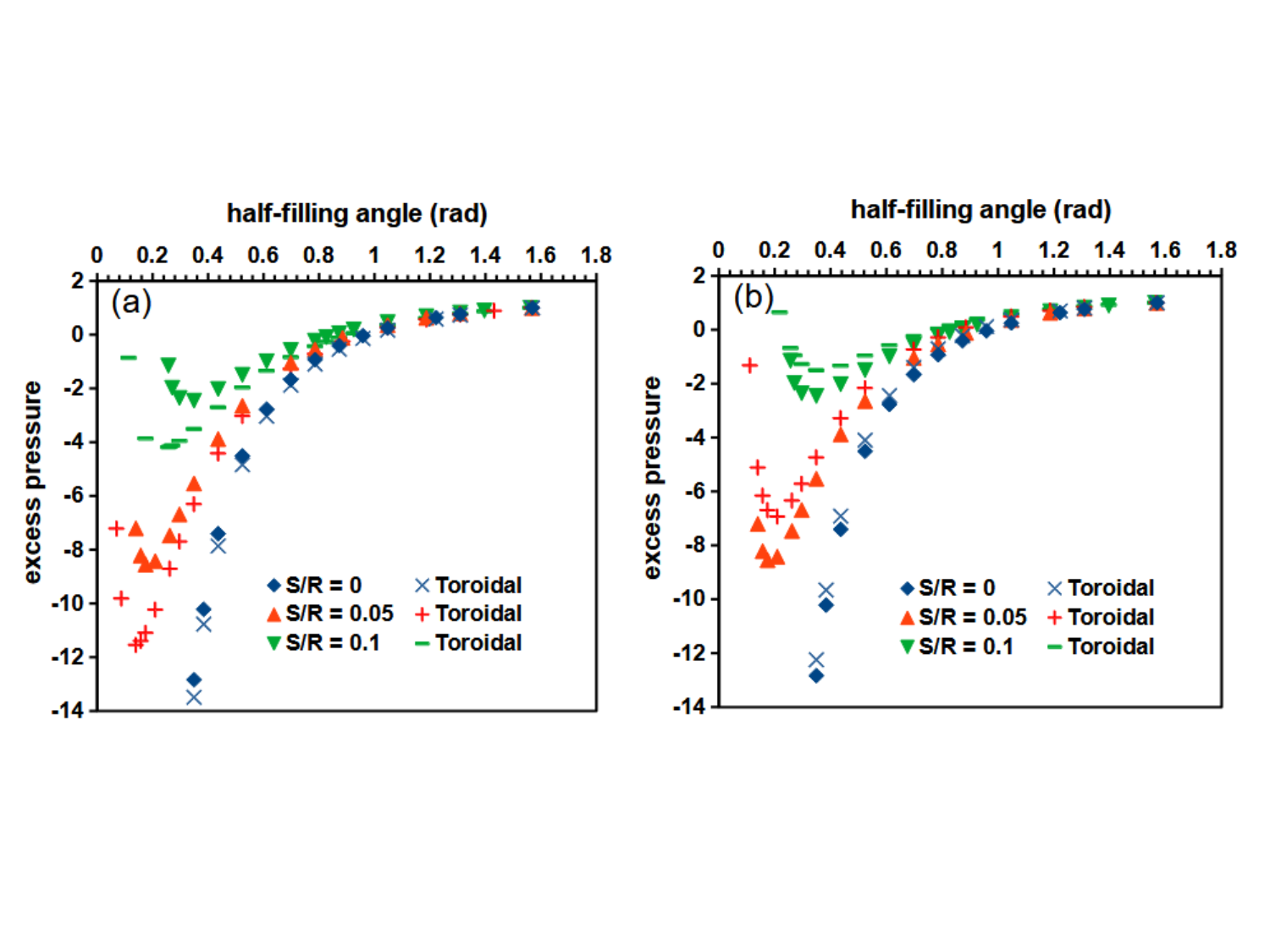}
\caption{Computed excess pressure $R \Delta p/\sigma$ ($= 2 H R$) in liquid bridge versus the half filling angle $\chi$ ($< \pi / 2$) for various $S/R$ , with comparison to the toroidal model based on (a) ${\widetilde H}_{\beta}$  and  (b) ${\widetilde H}_{0}$   }
\end{figure}

In fact, having a minimum corresponds to the existence of two roots for ${\widetilde H}_{0}$ $= 0$, 
which can easily be shown with the available analytical formulas (\ref{toroidalrhobeta}) 
because ${\widetilde H}_{0}$ $= 0$ is equivalent to 
\[
   (1 + S/R)[2 - \sin (\chi + \theta)] - \cos \chi + \sin \theta) = 0
   \quad  ,
\]
namely,
\[
\begin{array}{lllll}
   [(1 + S/R)^2 + 4 + 4 (1 + S/R)\sin \theta] \sin^2 \chi - \\ \\
   - 2 [2 (1 + S/R) + \sin \theta](1 + S/R) \cos \theta \sin \chi + \\ \\
   + (S/R) (2 + S/R) (4 - \sin^2 \theta) = 0  \\
\end{array}
   \quad  ,
\]
which would generally have two distinct roots.  But at $S = 0$,  we encounter a “cusp” singularity 
where the two roots degenerate to a single root
\[
   \sin \chi = \frac{2 (2 - \sin \theta) \cos \theta}{5 + 4 \sin \theta}
   \quad  .
\]
And for $\theta = 0$, the degenerated root becomes $\sin \chi = 4/5  = 0.8$ (at $S = 0$). 

From a physical point of view, the mean curvature based on the toroidal approximation ${\widetilde H}_0$ 
in (\ref{toroidalHrange}) displays a summation of two opposing terms $1 / \beta - 1 / \rho$ (with the neck radius $\beta \ge 0$
and the radius of curvature of bridge meridional profile $\rho \ge 0$ for concave meridian, as seen in figure 1 ).
For the excess pressure to become negative in the liquid bridge, we must have $\rho < \beta$. 
However, at a given half-filling angle $\chi$, increasing the half separation distance $S$ would 
lead to shrinking $\beta$ while enlarging $\rho$, namely, reducing the magnitude of the negative excess pressure 
consistently shown in figure 5. 
When $\chi \to \pi / 2$, $\beta \to r_c$ while $\rho \to \infty$ as the bridge approaching a cylindrical shape indicates.
With reducing $\chi$, both $\beta$ and $\rho$ decrease but with different behavior.
According to the formulas in (\ref{toroidalrhobeta}), we have $d\rho /d\chi > 0$ in the interval of $0 < \chi < \pi /2$ 
with minimum $\rho \propto S$ as $\chi \to 0$.  Therefore, $\rho \to 0$ can only happen when $S = 0$,  
while  $\beta$ can become $0$ at a nonzero $\chi$ where $\rho > 0$ unless $S = 0$.  
Hence the curve of ${\widetilde H}_0  = 1/\beta – 1/\rho$ versus $\chi$ would turn upward with reducing $\chi$ as long as $S \ne 0$ .

Thus, the excess pressure curves of $2 H R$ versus $\chi$ would have negative peaks, 
or minimum values, depending on the value of $S/R$, except for the case of $S = 0$ to decrease indefinitely 
with diminishing $\chi$.  Such minimum values of $2 H R$ correspond to the minimum “relative humidity” $h$ 
for capillary condensation to occur at given values of $S/R$ as can be calculated according to (\ref{kelvin}).  
A list of computed data for the case of water is provided in table 2 for $R = 25$ nm (corresponding to $50$ nm diameter nanoparticles) 
at $T = 358.15 K$ (i.e., $85 ^o$C) as a relevant situation for water-vapor assisted sintering of silver nanoparticle inks 
in printed electronics (cf. Bourassa et al. \cite{bourassa19}).  Hence at $85 ^o$C and $85\%$ RH—a typical condition 
in a humidity chamber for environmental test of printed electronics robustness—one would expect water vapor condensation 
to happen when $2 H R < -10$ with the mean radius of curvature $1/|2 H| < 0.1 R = 2.5$ nm in any $2 S < 2.5$ nm, 
possibly approaching the applicability limit of the Kelvin equation \cite{fisher81}. 

\begin{table}
  \begin{center}
\caption{Computed values of $\beta /R$, $F$, $V$ and $h$ corresponding to $\chi = \chi_{peak}$ for the negative peak of excess pressure $2HR$ at various specified $S/R$ for $R = 25$ nm and $T = 358.15 ^o$K (with $\theta = 0$)}  
  \begin{tabular}{lcccccc}
  \hline
    $S / R$ & $\chi_{peak} \, \mathrm{(deg)}$ & $2 H R$ & $\beta / R$ & $F$ & $V$ & $h$ \\[3pt]
  \hline    
     $0.001$ & 0.6 & -861.348 & 0.009387 & 0.094673 & $1.421 \times 10^{-7}$ & $4.6 \times 10^{-7}$ \\
     $0.002$ & 0.95 & -411.927 & 0.01437 & 0.11379 & $6.797 \times 10^{-7}$ & $9.3 \times 10^{-5}$ \\
     $0.005$ & 1.85 & -151.402 & 0.02656 & 0.15987 & $6.019 \times 10^{-6}$ & 0.0769 \\
     $0.01$ & 3.1 & -68.8228 & 0.04222 & 0.20712 & $3.160 \times 10^{-5}$ & 0.312 \\
     $0.02$ & 5.1 & -29.8976 & 0.06427 & 0.25206 & 0.0001541 & 0.603 \\
     $0.05$ & 10.5 & -8.58884 & 0.1170 & 0.35165 & 0.001398 & 0.865 \\
     $0.1$ & 19 & -2.45908 & 0.1913 & 0.47264 & 0.008133 & 0.959 \\
     $0.15$ & 26.5 & -0.71226 & 0.2481 & 0.53999 & 0.02167 & 0.988 \\
     $0.195$ & 33.5 & -0.007148 & 0.3054 & 0.61145 & 0.04387 & 0.99988 \\              
  \hline
  \end{tabular}
  \end{center}
\end{table}

If the particle size is increased to $R \sim 1$ $\mu$m, as those in the silty clay soil, for capillary condensation to occur at $0.95$, $0.85$, $0.75$ RH and ambient temperature (e.g., $25 ^o$C) the value of $2 H R$ would need to be $\sim -43.79$, $-138.74$, $-245.60$ corresponding to the mean radius of curvature $1/|2 H| \sim 23$, $7$, $4$ nm in a very small gap between particles (i.e., with $S < 13$, $6$, $4$ nm).  Apparently, the length scale for significant capillary condensation of water in clay soil is on the order of a few nanometers, barely within the validity limit of capillary theory \cite{giovambattista16, elliott21}. 

Noteworthy here from table 2 is the fact that capillary forces $F$ corresponding to the peak excess pressure decrease with reducing $S$, while the potential for capillary condensation increases with lowering $h$, as long as $S$ is nonzero.  Only when $S = 0$ could $F$ increase with lowering $h$ for contacting spheres. 

\subsubsection{Capillary forces}
Formation of a pendular liquid bridge between spheres would induce capillary forces ($F_{cap} \propto \sigma R$ on each sphere) with subsequent capillary cohesive strength—forces across the unit cross-sectional area—($\propto \sigma / R$ for various packing configurations of spheres) increasing as the size of spheres decreases.  Therefore, evaluating the value of capillary forces $F$ in (\ref{Fcap}) is of practical importance. 

Figure 6 shows curves of normalized capillary forces $F$ in (\ref{Fcap}) versus half filling angle $\chi$ for $S/R = 0$, $0.05$, and $0.1$.  Like that in figure 5, the curve of $S = 0$ is monotonic and apparently linear, quite different from those of nonzero $S$.  In contrast, with a nonzero $S$ the magnitude of $F$ would sharply decrease with reducing $\chi$ (e.g., for $\chi < 50^o$). The comparison between the toroidal values of ${\widetilde F}$ based on ${\widetilde H}_{\beta}$ and ${\widetilde H}_0$ (at $z = 0$) indicates that ${\widetilde H}_{\beta}$-based values over-estimates while ${\widetilde H}_0$-based ones generally under-estimates the capillary forces though as a relatively better approximation.  In general, the toroidal approximation for capillary forces does not seem too bad given its mathematical convenience with rigorously derived closed-form formulas. 

\begin{figure}[ht] \label{fig6}
\includegraphics[clip=true,scale=0.65,viewport=10 80 720 440]{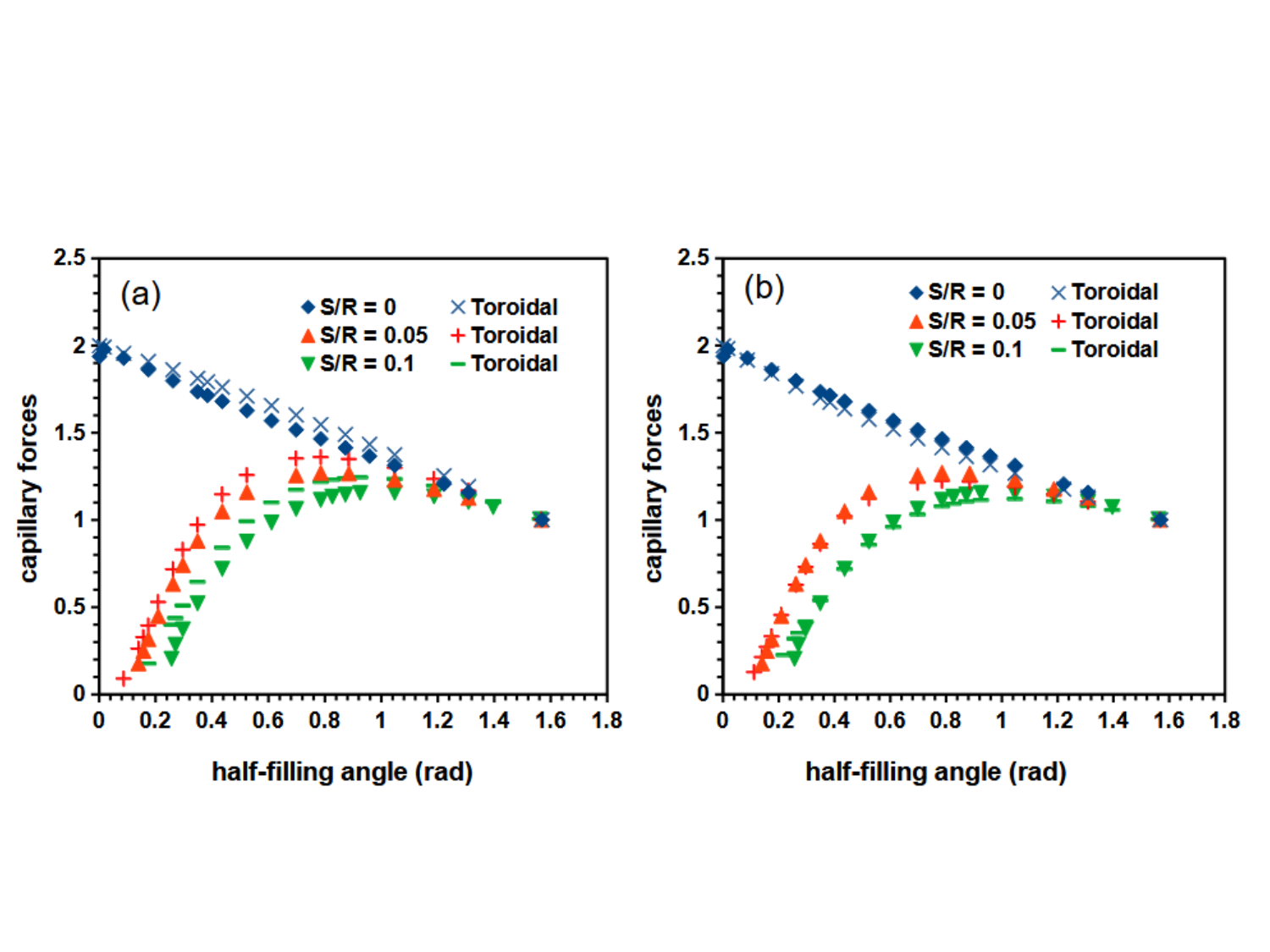}
\caption{As figure 5 but for capillary forces F computed according to (\ref{Fcap})   }
\end{figure}

Corresponding to figure 5 with the qualitative difference between the excess pressure $2 H R$ curve of $S = 0$ and those with nonzero $S$, the curve of capillary force for $S = 0$ also exhibits different characteristics from those of nonzero $S$.  With reducing $\chi$ the value of neck radius $\beta$  would shrink and, as a consequence, $F$ would also be expected to decrease with shrinking $\beta$ in view of the expression in (\ref{Fcap}) for $z = 0$.  Only when $S = 0$ would $F$ approach a finite value as $\chi$  and $\beta$ diminish where the value of $(1 – \beta H R)$ increases indefinitely (with $H R \to - \infty$ ).  Thus, the case of $S = 0$ has its special place in studying the capillary effects of pendular liquid bridges.
With the toroidal approximation in terms of ${\widetilde H}_0$ $= (1/\beta - 1/\rho)/2$ for $S = 0$, 
we would have ${\widetilde F} \to \beta^2 / (R \rho) \to 2$ at the limit of $\chi \to 0$ (and $\beta \to 0$, $\rho \to 0$), 
which becomes the same as $F$ (as $\chi \to 0$) from 
the Young-Laplace solution. 

As shown by Lian and Seville \cite{lian16} , the $S/R = 0$ curve can be fitted into a simple formula (in the present nomenclature)
\begin{equation}
   F = 2 - 0.66 \chi^{0.92}  \quad  \text{(with $\chi$ in radians) }
   \quad  ,
  \label{fittedF}
\end{equation}
which is quite accurate (for $\chi > \pi / 180$ with relative errors $< 0.8\%$).  However, such an exceptional success might be just fortuitous.  Several other curve-fitting formulas of Lian-Seville \cite{lian16} do not seem as accurate as desirable under close examination. 

\subsubsection{Formulas based on curve-fitting}
For convenient general usage, Lian and Seville \cite{lian16} provided some “closed-form” formulas obtained by curve-fitting numerically computed results, among which there is an exceptionally simple relationship between (normalized) liquid bridge volume and half-filling angle $\chi (< \pi / 3)$ for $\theta = 0$ and $S = 0$
\begin{equation}
   V = 0.156131 \chi^{3.73}  \quad  \text{(with $\chi$ in radians) }
   \quad  ,
  \label{fittedV}
\end{equation}
where $0.156131 = 0.654 \times 3 / (4 \pi)$ comes from the present definition of $V$ in (\ref{Vbridge}) which differs from that of Lian and Seville \cite{lian16} by a factor of $(4 \pi / 3)$.  However, (\ref{fittedV}) does not seem very accurate beyond $\chi = 45^o$ (or $\pi / 4$ radians) as shown in table 3, e.g., the error at $\chi = \pi / 4$ is $> 17\%$ and at $\chi = \pi / 3$ $> 28\%$.  
In contrast, the curve-fitting formulas in terms of polynomials (\ref{polynFittedV}) 
can be quite accurate for $\chi > \pi / 18$ radians (or $10^o$) with error $< 6\%$:
\begin{subeqnarray}
\label{polynFittedV}
   V = 0.139205 \chi^3 - 0.0259706 \chi^2 + 0.0017335 \chi \, \, , \, \mathrm{ for } \, \pi / 18 < \chi \le \pi / 3 \quad \slabel{polynFittedV0}\\
   V = 0.064212 \chi^4 - 0.054286 \chi^3 + 0.18199 \chi^2 - 0.0919611 \chi + 0.014827 \,  , \mathrm{ for } \, \chi > \pi / 3  \slabel{polynFittedV1} 
   \quad ,
\end{subeqnarray}
with $\chi$ in radians.
 Interestingly, 
the analytical formula from the toroidal model (\ref{toroidalVbridge}) seems to be quite accurate 
for calculating $V$ for  $\chi < \pi / 12$  (cf. Table 3 for $\chi < 15^o$) with errors $< 1.26\%$. 
In fact, the calculated values of liquid bridge volume using (\ref{toroidalVbridge}) do not deviate from the 
Young-Laplace values by more than $5\%$ for the entire interval of $0 < \chi < \pi/2$.

\begin{table}
  \begin{center}
\caption{Comparison among various curve-fitting formulas against computed values based on the Young-Laplace equation and those from the toroidal approximation (in parentheses) for $V$ at various $\chi$ with $S = 0$ and $\theta = 0$}  
  \begin{tabular}{lccccc}
  \hline
    $\chi \, \mathrm{(deg)}$ & $V$ & Eq. (\ref{fittedV}) & Eq. (\ref{polynFittedV0}) & Eq. (\ref{polynFittedV1}) \\[3pt]
  \hline    
     $5$ & $1.901 \times 10^{-5}$ ($1.905 \times 10^{-5}$) & $1.749 \times 10^{-5}$ & $4.614 \times 10^{-5}$ & 0.008155 \\
     $10$ & $2.674 \times 10^{-4}$ ($2.691 \times 10^{-4}$) & $2.321 \times 10^{-4}$ & $2.525 \times 10^{-4}$ & 0.004091 \\
     $15$ & 0.001197 (0.001212) & 0.001053 & 0.001175 & 0.002553 \\
     $30$ & 0.01376 (0.01419) & 0.01397 & 0.01379 & 0.01360 \\
     $45$ & 0.05282 (0.05523) & 0.06341 & 0.05287 & 0.05299 \\                             
     $60$ & 0.1333 (0.1397) & 0.1854 & 0.1334 & 0.1330 \\
     $75$ & 0.2731 (0.2823) & 0.4263 & 0.2704 & 0.2731 \\
     $89.9$ & 0.49811 (0.49826) & 0.8379 & 0.4772 & 0.4981 \\          
  \hline
  \end{tabular}
  \end{center}
\end{table}

Another curved-fitting formula of Lian and Seville \cite{lian16} is the liquid bridge neck radius as a monotonic function of half filling angle for $S = 0$ and $\theta = 0$, (\ref{fittedbeta0}), 
which does not seem very accurate with more than $10\%$ relative errors for $\chi > \pi /3$ ($> 15\%$ as $\chi$ approaches $\pi /2$).  
Again, a much more accurate formula can be obtained by a polynomial curve fitting, such as (\ref{fittedbeta1}), with relative errors $< 1\%$ for $\chi > \pi / 18$ ($1.65\%$ at $\chi = \pi / 36$ ):  
\begin{subeqnarray}
\label{fittedbeta}
   \beta (\chi) /R = 0.7634 \chi^{0.9179}  \, \, , \, \mathrm{ for } \, \, \chi < \pi / 6 \quad \quad \slabel{fittedbeta0}\\
   \beta (\chi) /R = 0.1127 \chi^3 - 0.3921 \chi^2 + 0.97503 \chi  \, \, , \, \mathrm{ for } \, \chi > \pi / 36 \quad \slabel{fittedbeta1} 
   \quad  ,
\end{subeqnarray}
with $\chi$ in radians. 

Substituting their power-law fitted formulas for (\ref{fittedbeta0}) and (\ref{fittedF}) into (\ref{Fcap}) for $z = 0$, Lian and Seville \cite{lian16} derived an approximate equation for the mean curvature (with $S = 0$ and $\theta = 0$), which is not very good outside the interval of $\pi / 180 < \chi <\pi / 4$.  If the formula (\ref{fittedF}) for $F(\chi)$ is used with (\ref{Fcap}) for $z = z_c$, without involving $\beta(\chi)$, the obtained equation
\begin{equation}
   2 H R = 2 \left(1 - \frac{1 - 0.33 \chi^{0.92}}{\sin^2 \chi} \right)
   \quad  ,
  \label{fittedH}
\end{equation}
with $\chi$ in radians, can be a much more accurate formula (with relative errors within a few percent for $\pi / 1800 < \chi < \pi / 2$) for calculating the mean curvature and excess pressure.

\subsection{Cases of nonzero contact angle $\theta$}
When the liquid is only partially wetting the solid spheres (i.e., $\theta \ne 0$, with $0 < \theta < \pi /2$), the range of valid half-filling angle shrinks to $0 < \chi < (\pi /2 \, – \, \theta$) for a liquid bridge with concave meridian.  As shown in figure 7, the qualitative behavior of normalized excess pressure ($= 2 H R$) versus half-filling angle $\chi$ for nonzero $\theta$ remains the same as that for $\theta = 0$.  However, the magnitude of negative peak value as well as the deviation of toroidal ${\widetilde H}_{\beta}$ from $H$ tends to decrease with increasing $\theta$ (in fact, the difference between the ${\widetilde H}_{\beta}$ curve and $H$ curve for the case of $\theta = 75^o$ would diminish to an almost unidentifiable level in the plot). 

\begin{figure}[ht] \label{fig7}
\includegraphics[clip=true,scale=0.65,viewport=10 80 720 440]{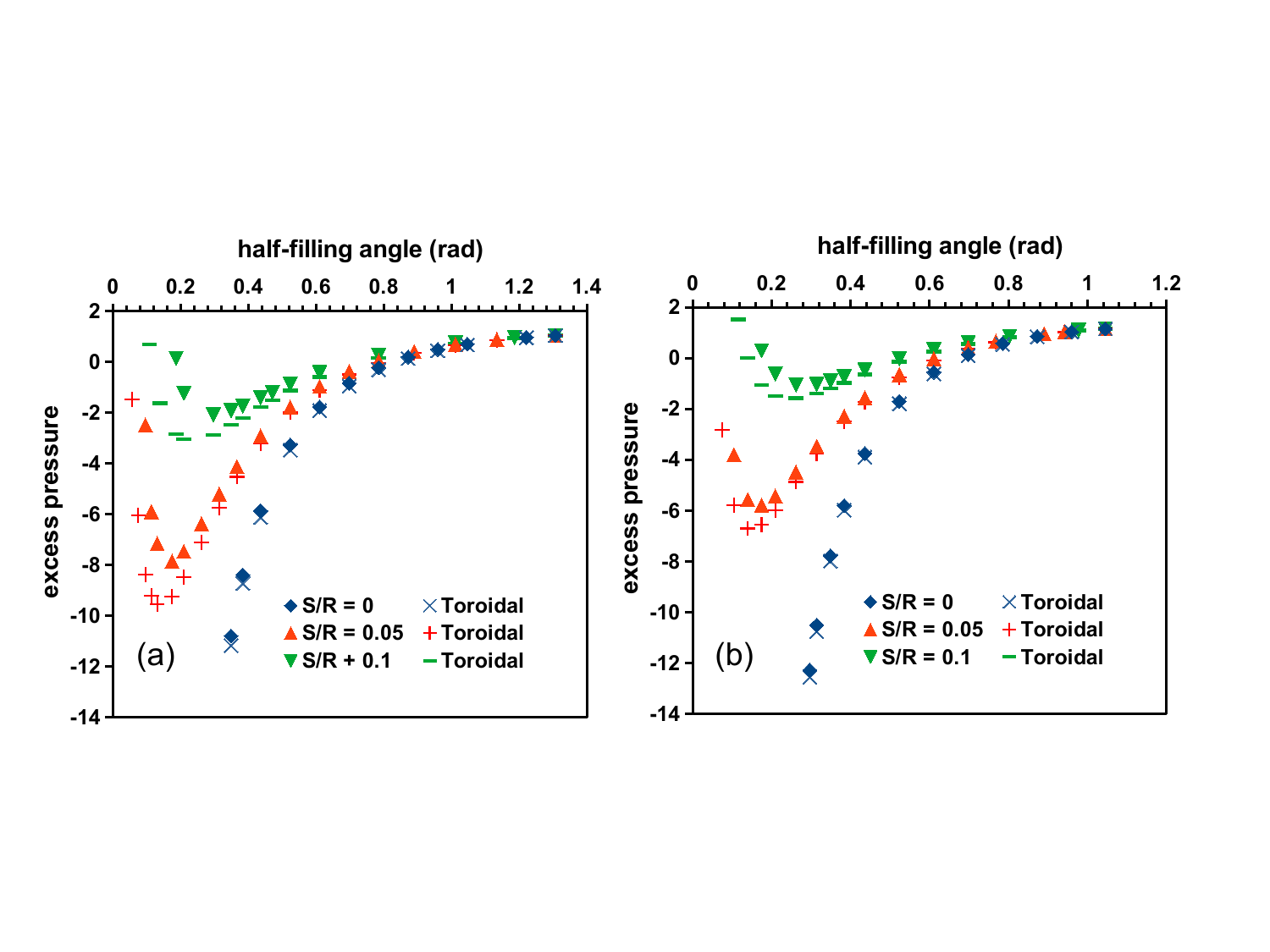}
\caption{As figure 5 but for the cases of nonzero contact angle (a) $\theta = 15^o$ ($\chi < 75^o$) and  (b) $\theta = 30^o$ ($\chi < 60^o$), with comparison to the toroidal model based on ${\widetilde H}_{\beta}$ }
\end{figure}

Again for $S = 0$, the magnitude of mean curvature can become indefinitely large with a diminishing half-filling angle corresponding to diminishing liquid bridge volume, as indicated by degenerated single root for $H = 0$ discussed in subsection 3.1.2.  Hence, a finite amount of liquid at a nonzero relative humidity would be expected to appear between two contacting spheres by virtue of capillary condensation.  For a nonzero $S$ whether a pendular liquid bridge can be maintained between two spheres depends on the value of $S/R$ and relative humidity, similar to qualitative trends shown in table 2.  The main effects of increasing $\theta$ are reducing the range of $\chi$ for liquid bridges with concave meridian to exist and the magnitude of peak negative mean curvature for a given value of $S/R$.  

Like the similarity of figures 5 and 7 for excess pressure, the curves of capillary forces for nonzero $\theta$  are quite similar to those in figure 6 too, and therefore unnecessary to be shown here. 

\section{Summary}
With a systematic study based on mathematical analysis, several less-recognized behaviors of liquid bridges between spheres are revealed in the present work.  Compared with the Young-Laplace solution, the accuracy of toroidal and elliptic approximations is examined with mathematical explanations.  At a given relative humidity, the separation distance between spheres can play an important role in pendular ring formation due to capillary condensation and the resulting magnitude of subsequent capillary forces, as explained with an in-depth analysis.  

With its relatively simple analytical formulas, the toroidal approximation can provide valuable mathematical insights at least in a qualitative sense, and be reasonably accurate for most practical cases (especially with diminishing separation distance).  To improve physical consistency, a new formula for calculating toroidal mean curvature is derived, differing from the commonly used formulas by previous authors.  Although using the elliptic meridional profile generally offers improved approximations from the toroidal model, the complexity of its analytical formulas would limit the convenience for practical applications.  

The present study also shows, with a few examples, that curve-fitting formulas cannot be perfect by their approximative nature and would always leave room for improvements. Therefore, care should be taken by checking their applicability conditions to avoid undesirable errors in serious applications.   


\section*{Funding}
None

\section*{Ethics declarations}
\subsection*{Conflict of interests}
The author has no conflicts to declare

\end{document}